\documentclass[twocolumn,superscriptaddress,showpacs]{revtex4-1}
\usepackage{amsmath,amssymb}
\usepackage[utf8x]{inputenc}
\usepackage[english]{babel}
\usepackage{graphicx,epsfig}
\usepackage{hyperref}
\usepackage{txfonts}

\begin{document}

\title{Modeling the metastable dynamics of correlated structures}
\author{Alexey M. Shakirov}
\email[Correspondence should be addressed to ]{a.shakirov@rqc.ru}
\affiliation{Department of Physics, Lomonosov Moscow State University, 119992 Moscow, Russia}
\affiliation{Russian Quantum Center, 143025 Skolkovo, Moscow region, Russia}
\author{Sergey V. Tsibulsky}
\affiliation{Department of Physics, Lomonosov Moscow State University, 119992 Moscow, Russia}
\author{Andrey E. Antipov}
\affiliation{University of Michigan, Ann Arbor, MI 48109, U.S.A.}
\author{Yulia E. Shchadilova}
\affiliation{Russian Quantum Center, 143025 Skolkovo, Moscow region, Russia}
\author{Alexey N. Rubtsov}
\affiliation{Department of Physics, Lomonosov Moscow State University, 119992 Moscow, Russia}
\affiliation{Russian Quantum Center, 143025 Skolkovo, Moscow region, Russia}

\begin{abstract}
Metastable quantum dynamics of an asymmetric triangular cluster that is coupled to a reservoir is investigated. The dynamics is governed by bath-mediated transitions, which in part require a thermal activation process. The decay rate is controlled by tuning the excitation spectrum of the frustrated cluster. We use the master equation approach and construct transition operators in terms of many-body states. We analyze dynamics of observables and reveal metastability of an excited state and of a magnetically polarized ground state.
\end{abstract}

\maketitle

\section*{Introduction}

Modern electronics relies on the ability of devices to switch internal states in a controllable manner. With the recent advantages in manufacturing complex nanoscale systems with engineered wavefunctions and energy surfaces, prediction of characteristic timescales has become an important and challenging task for a theory \cite{Capasso1990, Nitzan2003}. In particular, systems such as metal/oxide/metal memory cells \cite{Yang2008}, spintronic devices \cite{Sinova2012}, including quantum dots of single atoms and molecules on metallic surfaces \cite{Meier2008, Loth2010, Shchadilova2013}, and organic solar cells \cite{Gunes2007} are of a quantum nature and exhibit a complex energy landscape, resulting in a dynamics with multiple characteristic timescales.

In general, multiscale transition dynamics appear in any system with a complex energy surface and are enhanced when the system interacts with an external reservoir.
These dynamics manifest themselves in the presence of metastable states, in which the system stays for a long time. The existence of multiscale transition dynamics is not limited to a particular class of compounds and is possible in a wide range of systems, including chemical reactions \cite{Fukui1982}, polarons in molecular junctions \cite{Galperin2005,Alexandrov2003}, and black holes \cite{Casadio2008}. In a classical picture, metastability occurs for any isolated local energy minima. To describe quantum objects, one should consider metastable states as those from which a transition to lower energy states is prohibited by selection rules. The destruction of metastable states is possible via thermal activation to higher-lying energetic configurations or via tunneling. 

Nanoscale systems are of a quantum nature and the transitions occur between different multi-electron configurations. A natural way to observe metastability is by studying the relaxation dynamics in the presence of a weak coupling to the environment. It is usually represented as a reservoir of non-interacting particles and is described by the Hamiltonian $H_{E}=\sum_{k}\varepsilon_{k}c^{\dag}_{k}c_{k}$ with states $k$ forming a band. This reservoir is coupled to the system by particle exchange reflected in a hybridization Hamiltonian $H_{I}=\sum_{\alpha k}(\gamma_{\alpha k}c^{\dag}_{\alpha}c_{k}+\gamma^{\ast}_{\alpha k}c^{\dag}_{k}c_{\alpha})$ where subscript $\alpha$ enumerates localized states of the system and coefficients $\gamma_{\alpha k}$ represent their hybridization amplitudes. It is thus of crucial importance to provide a microscopic description of the dynamics which takes into account the full many-particle nature of the eigenstates of the systems, decoherence effects, and charge/energy transfer due to the coupling with the surrounding environment.

Various involved computational approaches such as time-dependent numerical renormalization group \cite{Anders2005}, time-dependent density matrix renormalization group \cite{White1992, Schmitteckert2004, DiasdaSilva2008, Heidrich-Meisner2009}, time-dependent functional renormalization group \cite{Jakobs2007, Kennes2012, Kennes2013} and other methods based on quantum Monte Carlo algorithms \cite{Muhlbacher2008, Schiro2009, Werner2009} do not allow one to access the behavior of the system at large times. On the other hand, the reduced density matrix theory \cite{Nakajima1958, Zwanzig1960, Tokuyama1976, Shibata1977, Zwanzig2001, Jin2008} provides a feasible way to access long-time properties of an interacting open quantum system. An exact version of this formalism can be coupled with the numerical methods mentioned above to obtain long-time dynamics at certain limits \cite{Cohen2011, Cohen2013a}. A frequently used approximation is described by the Redfield equation \cite{Pollard1996, Datta1995} which has been used in many contexts including nuclear magnetic resonance \cite{Wangsness1953, Bloch1957, Redfield1957}, quantum optics \cite{Scully1997, Gardiner2004, Breuer2002} and tunneling through quantum dots \cite{Timm2008, Koller2010, Timm2011, Cohen2013}.

In the reduced density matrix theory, the environment acts as a reservoir of particles and energy for the system and is often assumed to be of Markovian nature. A Fermi golden rule approach \cite{Dirac1927, Alicki1977} allows one to write the rate equations for diagonal elements of the density matrix $\rho_{S}(t)=\mbox{tr}_{E}\rho(t)$. The method of full counting statistics \cite{Bagrets2003, Braggio2006, Esposito2009} builds upon it to allow for an efficient evaluation of time-dependent observables and correlation functions. This approach neglects quantum interference effects between different eigenstates. To take them into account, one can employ
the Lindblad formalism \cite{Gorini1976, Lindblad1976, Alicki2007}. For a range of non-interacting problems the standard choice of transition operators as local creation and annihilation operators is known to be valid \cite{Prosen2008, Dzhioev2011, Horstmann2013}. The Lindblad formalism is also widely used for correlated systems interacting with an empty bath. This is a typical setup 
in the field of quantum optics where driven quantum systems emit photons to the vacuum but does not receive them back \cite{Breuer2002}.
Much less is known on the applicability of the Lindblad approach for the description of the correlated quantum systems that can exchange particles with the bath. This situation is relevant for the description of nanoclusters, molecules, quantum dots, etc. interacting with a substrate.

In this report we consider the correlated quantum triangular cluster exchanging electrons with the fermionic reservoir. We construct an extension of the standard Lindblad one-particle formalism for correlated open quantum systems exchanging particles with the bath. Using the obtained formalism, we describe the physics of metastability in the relaxation process of the correlated system.

We follow the standard approach of constructing the master equation for the reduced density matrix by employing a perturbation theory in powers of $H_{I}$ and considering the terms up to the second order (so-called a sequential tunneling approximation \cite{Timm2008, Timm2011}). The main distinction of the approach from the standard schemes that the transitions operators are constructed as matrices in the space of many-particle states. The time evolution in a leading contribution involves single-electron exchange with a reservoir. At low temperatures this results in a metastable many-body state of the system. We found that the metastablity arises from the specific structure of many-body spectrum and the coupling to a bath via the one-particle exchange channel. We consider a cluster of three atoms or quantum dots with local Coulomb interaction, which has a rich multiplet structure \cite{Mitchell2010, Mitchell2013} 
due to the interplay between frustration and interaction. The parameters of the cluster can be changed by using different isotopologues of nanoclusters or molecules \cite{Kumagai2012, Frederiksen2014}, while the substrate acts as a reservoir. The quantum simulation using ultracold atoms in optical traps is accessible. One can change the geometry of clusters and an interaction parameters directly \cite{Zimmermann2011} and the environment is represented by a surrounding cloud of particles.

\section*{Results and Discussion}

We considered the relaxation dynamics of an open quantum system. It is described by the master equation
\begin{equation}
\dfrac{d}{dt}\rho_{S}(t)=-i[H_{S},\rho_{S}(t)]+\mathcal{L}\rho_{S}(t).
\label{eqn:master}
\end{equation}
The first Liouvillean term governs the unitary evolution of the system with the Hamiltonian $H_{S}[c^{\dag}_{\alpha},c_{\alpha}]$. The second one takes into account the influence of the environment. It is obtained by applying a sequential tunneling approximation as described in Appendix A.

We found that $\mathcal{L}$ splits into so called one-particle $\mathcal{L}^{\infty}$ and many-particle $\mathcal{L}^{m}$ contributions. The first one is represented in terms of single-particle operators $c^{\dag}_{\alpha}$, $c_{\alpha}$ as in the well-known Lindblad formalism approach \cite{Breuer2002}. The second one has an essentially many-particle nature and vanishes in the following cases: 1) infinite temperature $T=\infty$, 2) infinite chemical potential $\mu=-\infty$ or $\mu=+\infty$. This implies that the influence of correlations in the system on its relaxation dynamics comes into play only when both the temperature and the chemical potential of the reservoir are not infinite. The many-particle part $\mathcal{L}^{m}$ of the superoperator is responsible for the relaxation to the Gibbs state at large time scales with the temperature $T$ and chemical potential $\mu$ determined by the reservoir. This is also the case in a Fermi golden rule approach, but in contrast to it, master equation formalism describes the evolution of both diagonal and off-diagonal elements of density matrix $\rho_{S}$.

\begin{figure}[t]
\includegraphics[width=0.5\columnwidth]{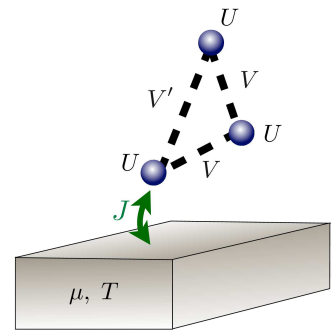}
\caption{Schematic representation of the triangular cluster with one site coupled to the reservoir with temperature $T$ and chemical potential $\mu$. It is modeled by the Fermi-Hubbard Hamiltonian in the half-filling regime with on-site interaction strength $U$ and hopping amplitudes $V=1$, $V'=2$. Coupling strength is characterized by a parameter $J$.}
\label{fgr:1}
\end{figure}

\begin{figure}[t]
\includegraphics[width=\columnwidth]{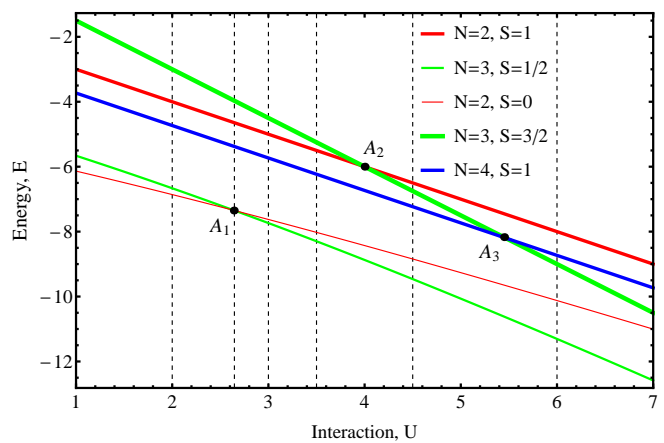}
\caption{Lowest energy terms as functions of the on-site interaction strength $U$. Different colors are used for terms with different occupation numbers, line thickness indicates their multiplicity $2S+1$. Level crossings $A_{1}$, $A_{2}$ and $A_{3}$ show points where tuning $U$ leads to the change in dynamics behavior.}
\label{fgr:2}
\end{figure}

We have modeled the multiscale dynamics in an asymmetric triangular quantum dot with local Coulomb interaction which is shown in Figure \ref{fgr:1}. The expression for the Hamiltonian of the system under investigation is
\begin{equation}\label{eq:Hamilt}
\begin{split}
&H_{S}=\sum_{\sigma=\uparrow,\downarrow}(Vc^{\dag}_{1\sigma}c_{2\sigma}+
Vc^{\dag}_{2\sigma}c_{3\sigma}+V'c^{\dag}_{3\sigma}c_{1\sigma}+\\
&+\mbox{h.c.})+U\sum_{i=1}^{3}(n_{i\uparrow}-\dfrac{1}{2})(n_{i\downarrow}-\dfrac{1}{2})\\
\end{split}
\end{equation}
Here $U$ is the on-site interaction strength, $V$, $V' = 2V$ are hopping amplitudes. This setup represents a minimal frustrated interacting model, where the frustration occurs due to an antiferromagnetic exchange interaction in the chain of odd number of nodes. The interplay between frustration and interaction gives rise to a nontrivial structure of the energy spectrum which can be tuned by parameters of the model. This allows us to probe different relaxation regimes by changing the single parameter - interaction strength $U$. We choose hopping amplitude $V$ as an energy scale of the system and break the $C_3$ rotational symmetry by setting $V'\neq V$. This separates dynamical effects originating from electronic correlations from those specified by geometry. States of the system are classified using integrals of motion -- the number of electrons $N$ in the cluster, the square of its spin momentum $S^{2}$ and the z-component $S_{z}$ of its spin momentum. Note that the chemical potential on the cluster is $\mu_{C}=U/2$. The triangular cluster lacks the electron-hole symmetry and its ground state does not necessarily correspond to $N=3$ electrons. The low energy part of the spectrum of the Hamiltonian (\ref{eq:Hamilt}) is shown in Figure \ref{fgr:2} as a function of $U$. At $U\approx 2.64$ the ground state changes from $N=2$ triplet state to $N=3$ duplet state. There are also level crossings of excited states, which may result in distinctly different decay behavior when the system is brought to the contact with the bath. We consider dynamics when the bath is weakly coupled to the cluster. The details of the formalism used for the description of the relaxation processes of the open system in the regime of strong correlations are presented in Appendix B.

The relaxation process is determined by the selection rules, $\Delta N=\pm1$, $\Delta S_{z}=\pm1/2$. These are imposed on transitions between many-body states because the elementary process generated by the interaction Hamiltonian is a transfer of one electron. The resulting metastability includes (i) slow mixing of degenerate ground states which are not coupled by direct transitions, and (ii) slow decay of non-ground states from which direct transitions to lower energy ones are not allowed. In both cases the relaxation happens through transitions to intermediate states lying higher in energy, and its rate $R_{1}\propto J\mbox{exp}(-\Delta E/T)$ depends on the temperature of the reservoir, and the effective coupling parameter $J$. The increase in $T$ causes the system to escape faster from the metastable state. This holds in the time region $t\lesssim V/J^{2}$ where the thermal activation is the only mechanism of metastability decay. For larger times the effects of next order perturbations come into play. They allow exchanges with the reservoir by two particles at a rate $R_{2}\propto J^{2}/V$. Transitions between states with the same occupation number and $\Delta S_{z}=\pm1$ act as another mechanism of the decay process and should be taken into account at low temperatures when the thermal activation rate is negligible.

\begin{figure}[t]
\includegraphics[width=\columnwidth]{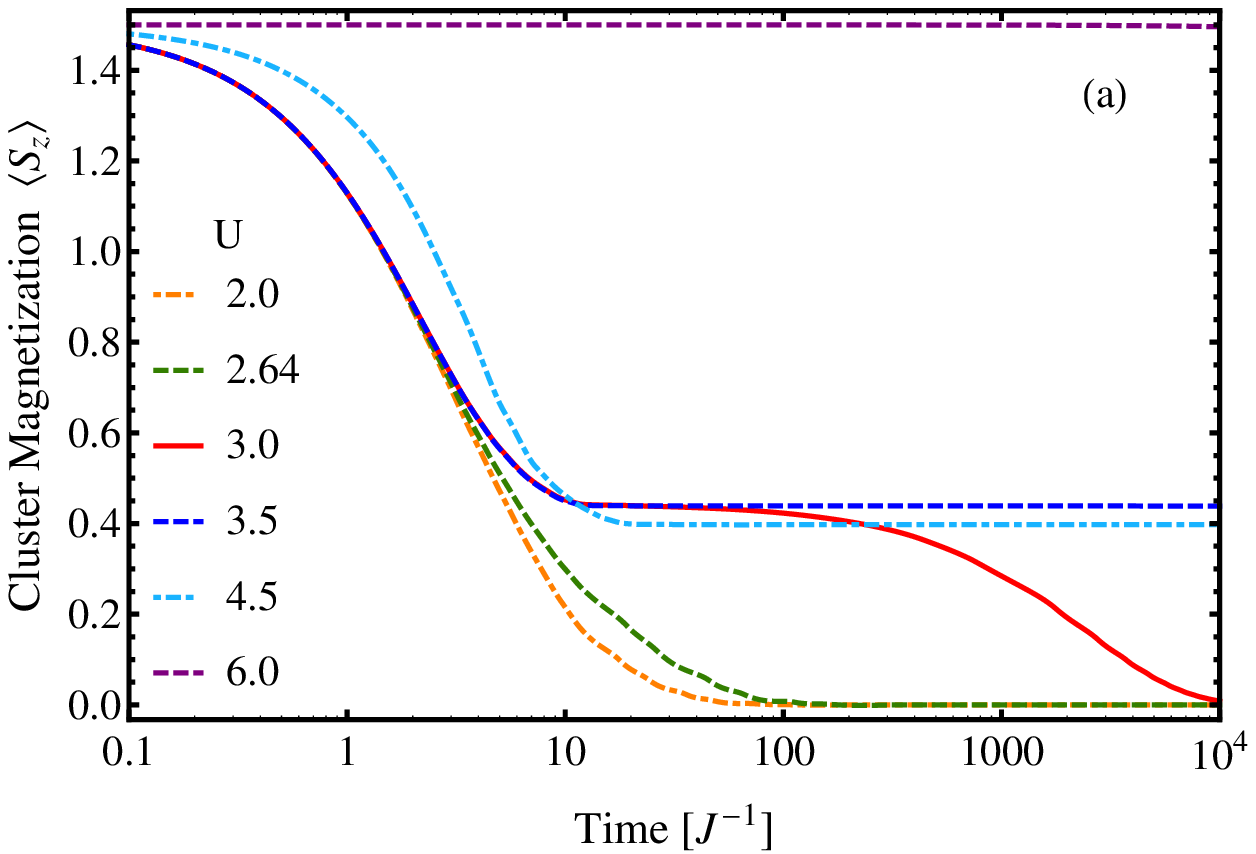}
\includegraphics[width=\columnwidth]{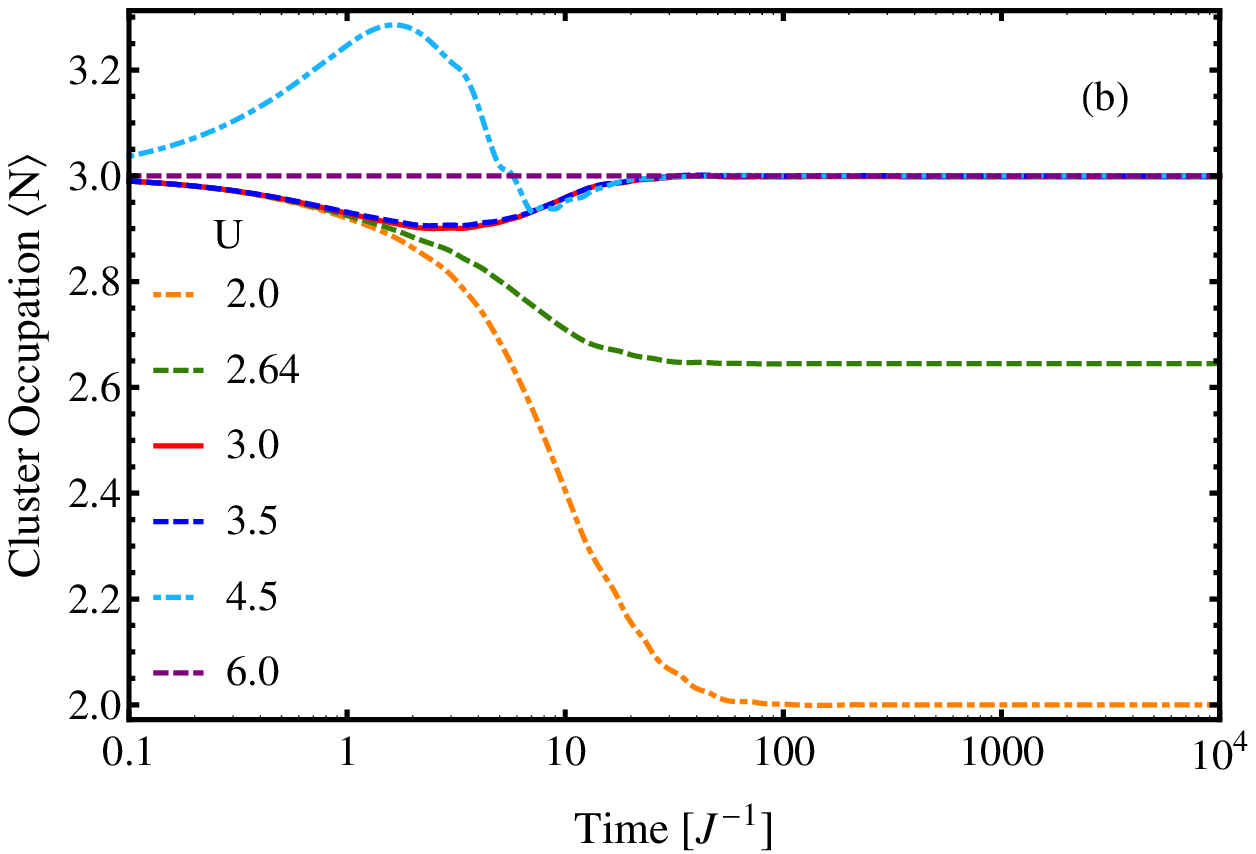}
\caption{The decay of a fully spin polarized state in the frustrated triangular cluster coupled to the reservoir at temperature $T=0.02$. The interaction parameter $U$ is tuned as indicated by different colors. Dynamics of (a) cluster magnetization $\langle S_{z}\rangle$ and (b) cluster occupation $\langle N\rangle$ is presented. One can see that tuning $U$ allows to control the relaxation process of the excited state of the cluster.}
\label{fgr:3}
\end{figure}

We demonstrate discussed features of relaxation in the decay of a fully spin polarized state $N=3$, $S=3/2$, $S_{z}=3/2$ at small temperature. It can be prepared by applying an external static magnetic field to the system. The behavior of dynamics depends on the interaction strength, so different values of $U$ are considered. They are taken from the regions delimited by level crossing points $A_{1}$, $A_{2}$ and $A_{3}$ as shown in Figure \ref{fgr:2}. The dynamics are manifested in the expectation values of observables $S_{z}(t)$ and $N(t)$ as shown in Figure \ref{fgr:3}. Their time dependence levels out at characteristic times $Jt\sim10-100$. For $U=2.0$ (lying below $A_{1}$) there is a transition channel from the initial state to the ground state, so that metastability does not appear. For $U=2.64$ (lying at $A_{1}$) the system relaxes to an equiprobable mixture of a singlet state and two doublet states with $\langle N\rangle=8/3$ and $\langle S_{z}\rangle=0$ at large times. This mixing becomes metastable in the region above $A_{1}$. The system moves to the state with $S_{z}=1/2$ of the ground doublet while the state with $S_{z}=-1/2$ can only be reached through an excitation to the singlet state. This process might be accelerated by tuning the system closer to $A_{1}$ -- compare $U=3.0$ and $U=3.5$. Passing through $A_{2}$ point changes the number of transition channels leading to the ground state from $2$ to $1$. The curve for $U=4.5$ levels out more slowly than the curve for $U=3.5$ because the intermediate state $N=2$, $S=1$ becomes higher in energy than the initial state. Increasing $U$ above $A_{3}$ point blocks the transition through the second intermediate state $N=4$, $S=1$ as well. For $U=6.0$ an initial fully spin polarized state becomes metastable with respect to the decay.

The existence of multiple crossing points ($A_1$, $A_2$ and $A_3$) in Figure \ref{fgr:2} upon the variation of onsite interaction strength $U$ of both ground and excited states reveals an important implication of the role of frustration in correlated fermionic systems. Not only it leads to a multitude of possible ground-state phases, but also to a rich manifold of metastability patterns upon the decay from excited states. The considered system can be seen as a constituting block for the anisotropic Hubbard model on a triangular lattice, which receives substantial interest being prototypical to the spin-liquid physics of the correlated charge-transfer organic salts \cite{Li2014}. Richness of the phase diagram is related to an interplay between frustration effects and the anisotropy of the triangular lattice; the latter is present in our consideration as $V\neq V'$. We therefore expect rich metastable relaxation dynamics patterns, controlled by the frustration of the lattice in other frustrated strongly correlated systems, which represents a substantial interest for further studies.

The presented approach requires operations with the density matrix and can be directly applied to the correlated cluster of only a small size, due to the exponential growth of the Hilbert space of the system. Certain increase of the system size can be achieved when the truncation of the many-body spectrum is possible. This is the case for systems where a limited number of transition channels works \cite{Reiter2012, Schuetz2013} or at small temperatures of the bath when only the lowest energy states are involved in the dynamics. The relevant part of the many-body spectrum can be efficiently obtained by the Lanczos method. For a construction of the many-body density matrix, there is a variety of established approximate methods of computational quantum chemistry of molecules, including Hartree-Fock method and its extensions \cite{Cramer2013}.

To summarize, we considered a correlated triangular cluster where the interplay between interaction and frustration results in level crossings which have an effect on the decay behaviour. We have shown that the dynamics of open correlated quantum systems requires an extensiton of the standard Lindblad one-particle formalism. Working with a many-body spectrum of the correlated triangular cluster, we were able to capture its non-trivial multi-scale relaxation dynamics to the thermal equilibrium as a result of weak coupling to the reservoir. We demonstrated the dynamics of the spin and charge degrees of freedom of the system and explained the arising metastability by selection rules which are imposed on transitions between many-body states. The multistage relaxation process involves the excitation to the states lying higher in energy. As a result, the time scale of dynamics becomes dependent on the temperature of the reservoir. The approach can be applied to a number of problems in quantum chemistry, nanoelectronics and quantum information technology.

\section*{Acknowledgements}
We are grateful to A. I. Lichtenstein, D. Zgid, G. Cohen, Ch. H{\"u}bner and J. P. F. LeBlanc for fruitful discussions. A financial support from Dynasty foundation and RFBR 14-02-0121914 is acknowledged. S.V.Ts. and A.E.A. are grateful to Russian Quantum Center for hospitality.

\section*{Appendix A}

In the master equation approach a weak hybridization of an open quantum system with the environment is treated in a perturbative manner. At the same time correlations are taken into account exactly by calculating many-particle eigenstates $|n\rangle$ and eigenenergies $E_{n}$ of the isolated system. Small reservoir excitations from the equilibrium Gibbs state $\rho_{B}$ normally decays fast on a resolved time-scale, so the Markov approximation holds. In the lowest non-vanishing order the resulting equation reads
\begin{equation}
\begin{split}
&\dfrac{d}{dt}\rho_{S}(t)=-i[H_{S},\rho_{S}(t)]-\int\limits_{0}^{\infty}dt'\mbox{tr}_{E}\\
&[H_{I},[e^{-i(H_{S}+H_{E})t'}H_{I}e^{i(H_{S}+H_{E})t'},
\rho_{S}(t)\rho_{E}]].\\
\end{split}
\label{eqn:start}
\end{equation}
Substituting $H_{E}$, $H_{I}$ in this expression is covered in details in Appendix B. For a practical purpose we discard the dependence on energy of the hybridization function $J_{\alpha\beta}(\varepsilon)=2\pi\sum_{k}
\gamma_{\alpha k}\gamma^{\ast}_{\beta k}\delta(\varepsilon-\varepsilon_{k})$ assuming that a wide band approximation is fullfilled. Transition operators are defined as follows
\begin{equation}
\begin{split}
&L_{\alpha}=\sum_{mn}f(\varepsilon_{nm})\langle m|c_{\alpha}|n\rangle|m\rangle\langle n|\\
&\bar{L}_{\alpha}=\sum_{mn}(1-f(\varepsilon_{nm}))\langle m|c_{\alpha}|n\rangle|m\rangle\langle n|,\\
\end{split}
\label{eqn:operators}
\end{equation}
where $f(\varepsilon)=(\mbox{exp}(\frac{\varepsilon-\mu}{T})+1)^{-1}$ is a Fermi distribution function. The master equation takes the form
\begin{equation}
\begin{split}
&\dfrac{d}{dt}\rho_{S}(t)=-i[H_{S},\rho_{S}(t)]+\frac{1}{2}\sum_{\alpha\beta}J_{\alpha\beta}
(c^{\dag}_{\alpha}\rho_{S}(t)L_{\beta}+\\
&+\bar{L}_{\beta}\rho_{S}(t)c^{\dag}_{\alpha}-
c^{\dag}_{\alpha}\bar{L}_{\beta}\rho_{S}(t)-
\rho_{S}(t)L_{\beta}c^{\dag}_{\alpha})+\mbox{h.c.}\\
\end{split}
\label{eqn:final}
\end{equation}
There is a relation of $L_{\alpha}$, $\bar{L}_{\alpha}$ to the Lindblad formalism approach employing operators $c^{\dag}_{\alpha}$ and $c_{\alpha}$ in the master equation. The following correspondence was found: 1) $L_{\alpha}=\bar{L}_{\alpha}=\frac{1}{2}c_{\alpha}$ at $T=\infty$, 2) $L_{\alpha}=0$, $\bar{L}_{\alpha}=c_{\alpha}$ at $\mu=-\infty$, and 3) $L_{\alpha}=c_{\alpha}$, $\bar{L}_{\alpha}=0$ at $\mu=+\infty$. Though for concreteness the dynamics of fermions is considered in the paper, the developed formalism can be applied to both kinds of particle statistics.

\section*{Appendix B}

We derive the master equation $\dot{\rho}_{S}(t)=-i[H_{S},\rho_{S}(t)]-\mathcal{L}\rho_{S}(t)$ for the reduced density matrix of an open quantum system. First we substitute $H_{E}=\sum_{k}\varepsilon_{k}c^{\dag}_{k}c_{k}$ and $H_{I}=\sum_{\alpha k}(\gamma_{\alpha k}c^{\dag}_{\alpha}c_{k}+\gamma^{\ast}_{\alpha k}c^{\dag}_{k}c_{\alpha})$ into
\begin{equation}
\begin{split}
&\dfrac{d}{dt}\rho_{S}(t)=-i[H_{S},\rho_{S}(t)]-\int\limits_{0}^{\infty}dt'\mbox{tr}_{E}\\
&[H_{I},[e^{-i(H_{S}+H_{E})t'}H_{I}e^{i(H_{S}+H_{E})t'},\rho_{S}(t)\rho_{E}]]\\
\end{split}
\label{eqn_initial}
\end{equation}
and clarify the action of operator exponents
\begin{equation}
\begin{split}
&e^{-iH_{E}t}c_{k}e^{iH_{E}t}=e^{i\varepsilon_{k}t}c_{k}\\
&e^{-iH_{S}t}c_{\alpha}e^{iH_{S}t}=\sum_{mn}\langle m|c_{\alpha}|n\rangle
e^{-i(E_{m}-E_{n})t}|m\rangle\langle n|\\
\end{split}
\label{A1}
\end{equation}
Here $|n\rangle$, $E_{n}$ are eigenstates and eigenergies of $H_{S}$ obtained by an exact diagonalization. We assume that $\rho_{E}$ is a Gibbs state and leave only terms with pairs of reservoir operators which do not vanish after taking partial trace. This transforms the non-Liouvillean part into
\begin{equation}
\begin{split}
&\mathcal{L}\rho_{S}(t)=\int\limits_{0}^{\infty}dt'\mbox{tr}_{E}\sum_{\substack{\alpha\beta k\\mn}}e^{i(E_{n}-E_{m}-\varepsilon_{k})t'}[\gamma_{\alpha k}c^{\dag}_{\alpha}c_{k},\\
&[\gamma^{\ast}_{\beta k}\langle m|c_{\beta}|n\rangle c^{\dag}_{k}|m\rangle\langle n|,\rho_{S}(t)\rho_{E}]]+\mbox{h.c.}\\
\end{split}
\label{A2}
\end{equation}
Taking time integral and partial trace leads to
\begin{equation}
\begin{split}
&\mathcal{L}\rho_{S}(t)=\pi\sum_{\substack{\alpha\beta k\\mn}}\delta(E_{n}-E_{m}-\varepsilon_{k})
\gamma_{\alpha k}\gamma^{\ast}_{\beta k}\langle m|c_{\beta}|n\rangle\\
&(\bar{f}_{k}c^{\dag}_{\alpha}|m\rangle\langle n|\rho_{S}(t)-
f_{k}c^{\dag}_{\alpha}\rho_{S}(t)|m\rangle\langle n|-\\
&-\bar{f}_{k}|m\rangle\langle n|\rho_{S}(t)c^{\dag}_{\alpha}+
f_{k}\rho_{S}(t)|m\rangle\langle n|c^{\dag}_{\alpha})+\mbox{h.c.}\\
\end{split}
\label{A5}
\end{equation}
Here $\mbox{tr}_{E}(c^{\dag}_{k}c_{k}\rho_{E})=f(\varepsilon_{k})=(\mbox{exp}(\frac{\varepsilon_{k}-\mu}{T})+1)^{-1}$ is a distribution function of particles in the reservoir and $\mbox{tr}_{E}(c_{k}c^{\dag}_{k}\rho_{E})=\bar{f}(\varepsilon_{k})=(1+\mbox{exp}(-\frac{\varepsilon_{k}-\mu}{T}))^{-1}$. We also introduce hybridization function $J_{\alpha\beta}(\varepsilon)=2\pi\sum_{k}\gamma_{\alpha k}
\gamma^{\ast}_{\beta k}\delta(\varepsilon-\varepsilon_{k})$ which allows to rewrite the last expression in the following form
\begin{equation}
\begin{split}
&\mathcal{L}\rho_{S}(t)=\frac{1}{2}\sum_{\substack{\alpha\beta\\mn}}J_{\alpha\beta}(\varepsilon_{nm})\langle m|c_{\beta}|n\rangle\\
&(\bar{f}(\varepsilon_{nm})c^{\dag}_{\alpha}|m\rangle\langle n|\rho_{S}(t)-
f(\varepsilon_{nm})c^{\dag}_{\alpha}\rho_{S}(t)|m\rangle\langle n|-\\
&-\bar{f}(\varepsilon_{nm})|m\rangle\langle n|\rho_{S}(t)c^{\dag}_{\alpha}+
f(\varepsilon_{nm})\rho_{S}(t)|m\rangle\langle n|c^{\dag}_{\alpha})+\mbox{h.c.}\\
\end{split}
\label{A6}
\end{equation}
We note that the matrix structure of $J_{\alpha\beta}(\varepsilon)$ is determined by a hybridization of the reservoir with different localized states. Its dependence on energy is ignored in the wide band approximation. Then we introduce two kinds of transition operators
\begin{equation}
\begin{split}
&L_{\alpha}=\sum_{mn}f(\varepsilon_{nm})\langle m|c_{\alpha}|n\rangle|m\rangle\langle n|\\
&\bar{L}_{\alpha}=\sum_{mn}\bar{f}(\varepsilon_{nm})\langle m|c_{\alpha}|n\rangle|m\rangle\langle n|\\
\end{split}
\label{A7}
\end{equation}
They allow to simplify (\ref{A6}) and finally write the master equation as
\begin{equation}
\begin{split}
&\dfrac{d}{dt}\rho_{S}(t)=-i[H_{S},\rho_{S}(t)]+\frac{1}{2}\sum_{\alpha\beta}J_{\alpha\beta}
(c^{\dag}_{\alpha}\rho_{S}(t)L_{\beta}+\\
&+\bar{L}_{\beta}\rho_{S}(t)c^{\dag}_{\alpha}-
c^{\dag}_{\alpha}\bar{L}_{\beta}\rho_{S}(t)-
\rho_{S}(t)L_{\beta}c^{\dag}_{\alpha})+\mbox{h.c.}\\
\end{split}
\label{A8}
\end{equation}
We note that this derivation is also applicable to bosons if one uses $f(\varepsilon_{k})=(\mbox{exp}(\frac{\varepsilon_{k}-\mu}{T})-1)^{-1}$ and $\bar{f}(\varepsilon_{k})=(1-\mbox{exp}(-\frac{\varepsilon_{k}-\mu}{T}))^{-1}$.


\begin{thebibliography}{10}
\expandafter\ifx\csname url\endcsname\relax
  \def\url#1{\texttt{#1}}\fi
\expandafter\ifx\csname urlprefix\endcsname\relax\def\urlprefix{URL }\fi
\providecommand{\bibinfo}[2]{#2}
\providecommand{\eprint}[2][]{\url{#2}}

\bibitem{Capasso1990}
\bibinfo{author}{Capasso, F.}
\newblock \emph{\bibinfo{title}{{Physics of Quantum Electron Devices.}}}
  (\bibinfo{publisher}{Springer-Verlag}, \bibinfo{city}{Berlin}, \bibinfo{year}{1990}).

\bibitem{Nitzan2003}
\bibinfo{author}{Nitzan, A.} \& \bibinfo{author}{Ratner, M.~A.}
\newblock \bibinfo{title}{{Electron transport in molecular wire junctions.}}
\newblock \emph{\bibinfo{journal}{Science}} \textbf{\bibinfo{volume}{300}},
  \bibinfo{pages}{1384--1389} (\bibinfo{year}{2003}).

\bibitem{Yang2008}
\bibinfo{author}{Yang, J.~J.} \emph{et~al.}
\newblock \bibinfo{title}{{Memristive switching mechanism for metal/oxide/metal
  nanodevices.}}
\newblock \emph{\bibinfo{journal}{Nat. Nanotechnol.}}
  \textbf{\bibinfo{volume}{3}}, \bibinfo{pages}{429--433}
  (\bibinfo{year}{2008}).

\bibitem{Sinova2012}
\bibinfo{author}{Sinova, J.} \& \bibinfo{author}{\v{Z}uti\'{c}, I.}
\newblock \bibinfo{title}{{New moves of the spintronics tango.}}
\newblock \emph{\bibinfo{journal}{Nat. Mater.}} \textbf{\bibinfo{volume}{11}},
  \bibinfo{pages}{368--371} (\bibinfo{year}{2012}).

\bibitem{Meier2008}
\bibinfo{author}{Meier, F.}, \bibinfo{author}{Zhou, L.},
  \bibinfo{author}{Wiebe, J.} \& \bibinfo{author}{Wiesendanger, R.}
\newblock \bibinfo{title}{{Revealing magnetic interactions from single-atom
  magnetization curves.}}
\newblock \emph{\bibinfo{journal}{Science}} \textbf{\bibinfo{volume}{320}},
  \bibinfo{pages}{82--6} (\bibinfo{year}{2008}).

\bibitem{Loth2010}
\bibinfo{author}{Loth, S.}, \bibinfo{author}{Etzkorn, M.},
  \bibinfo{author}{Lutz, C.~P.}, \bibinfo{author}{Eigler, D.~M.} \&
  \bibinfo{author}{Heinrich, A.~J.}
\newblock \bibinfo{title}{{Measurement of Fast Electron Atomic Resolution.}}
\newblock \emph{\bibinfo{journal}{Science}} \textbf{\bibinfo{volume}{329}},
  \bibinfo{pages}{1628--1630} (\bibinfo{year}{2010}).

\bibitem{Shchadilova2013}
\bibinfo{author}{Shchadilova, Y.~E.}, \bibinfo{author}{Tikhodeev, S.~G.},
  \bibinfo{author}{Paulsson, M.} \& \bibinfo{author}{Ueba, H.}
\newblock \bibinfo{title}{{Rotation of a Single Acetylene Molecule on Cu(001)
  by Tunneling Electrons in STM.}}
\newblock \emph{\bibinfo{journal}{Phys. Rev. Lett.}}
  \textbf{\bibinfo{volume}{111}}, \bibinfo{pages}{186102}
  (\bibinfo{year}{2013}).

\bibitem{Gunes2007}
\bibinfo{author}{G\"{u}nes, S.}, \bibinfo{author}{Neugebauer, H.} \&
  \bibinfo{author}{Sariciftci, N.~S.}
\newblock \bibinfo{title}{{Conjugated polymer-based organic solar cells.}}
\newblock \emph{\bibinfo{journal}{Chem. Rev.}} \textbf{\bibinfo{volume}{107}},
  \bibinfo{pages}{1324--1338} (\bibinfo{year}{2007}).

\bibitem{Fukui1982}
\bibinfo{author}{Fukui, K.}
\newblock \bibinfo{title}{{Role of frontier orbitals in chemical reactions.}}
\newblock \emph{\bibinfo{journal}{Science}} \textbf{\bibinfo{volume}{218}},
  \bibinfo{pages}{747--754} (\bibinfo{year}{1982}).

\bibitem{Galperin2005}
\bibinfo{author}{Galperin, M.}, \bibinfo{author}{Ratner, M.~A.} \&
  \bibinfo{author}{Nitzan, A.}
\newblock \bibinfo{title}{{Hysteresis, switching, and negative differential
  resistance in molecular junctions: a polaron model.}}
\newblock \emph{\bibinfo{journal}{Nano Lett.}} \textbf{\bibinfo{volume}{5}},
  \bibinfo{pages}{125--130} (\bibinfo{year}{2005}).

\bibitem{Alexandrov2003}
\bibinfo{author}{Alexandrov, A.} \& \bibinfo{author}{Bratkovsky, A.}
\newblock \bibinfo{title}{{Memory effect in a molecular quantum dot with strong
  electron-vibron interaction.}}
\newblock \emph{\bibinfo{journal}{Phys. Rev. B}} \textbf{\bibinfo{volume}{67}},
  \bibinfo{pages}{235312} (\bibinfo{year}{2003}).

\bibitem{Casadio2008}
\bibinfo{author}{Casadio, R.} \& \bibinfo{author}{Nicolini, P.}
\newblock \bibinfo{title}{{The decay-time of non-commutative micro-black
  holes.}}
\newblock \emph{\bibinfo{journal}{J. High Energy Phys.}}
  \textbf{\bibinfo{volume}{2008}}, \bibinfo{pages}{072--072}
  (\bibinfo{year}{2008}).

\bibitem{Anders2005}
\bibinfo{author}{Anders, F.~B.} \& \bibinfo{author}{Schiller, A.}
\newblock \bibinfo{title}{{Real-Time Dynamics in Quantum-Impurity Systems: A
  Time-Dependent Numerical Renormalization-Group Approach.}}
\newblock \emph{\bibinfo{journal}{Phys. Rev. Lett.}}
  \textbf{\bibinfo{volume}{95}}, \bibinfo{pages}{196801}
  (\bibinfo{year}{2005}).

\bibitem{White1992}
\bibinfo{author}{White, S.~R.}
\newblock \bibinfo{title}{{Density Matrix Formulation for Quantum
  Renormalization Groups.}}
\newblock \emph{\bibinfo{journal}{Phys. Rev. Lett.}}
  \textbf{\bibinfo{volume}{69}}, \bibinfo{pages}{2863--2866}
  (\bibinfo{year}{1992}).

\bibitem{Schmitteckert2004}
\bibinfo{author}{Schmitteckert, P.}
\newblock \bibinfo{title}{{Nonequilibrium electron transport using the density
  matrix renormalization group method.}}
\newblock \emph{\bibinfo{journal}{Phys. Rev. B}} \textbf{\bibinfo{volume}{70}},
  \bibinfo{pages}{121302} (\bibinfo{year}{2004}).

\bibitem{DiasdaSilva2008}
\bibinfo{author}{da~Silva, L. G. G. V.~D.} \emph{et~al.}
\newblock \bibinfo{title}{{Transport properties and Kondo correlations in
  nanostructures: Time-dependent DMRG method applied to quantum dots coupled to
  Wilson chains.}}
\newblock \emph{\bibinfo{journal}{Phys. Rev. B}} \textbf{\bibinfo{volume}{78}},
  \bibinfo{pages}{195317} (\bibinfo{year}{2008}).

\bibitem{Heidrich-Meisner2009}
\bibinfo{author}{Heidrich-Meisner, F.}, \bibinfo{author}{Feiguin, A.~E.} \&
  \bibinfo{author}{Dagotto, E.}
\newblock \bibinfo{title}{{Real-time simulations of nonequilibrium transport in
  the single-impurity Anderson model.}}
\newblock \emph{\bibinfo{journal}{Phys. Rev. B}} \textbf{\bibinfo{volume}{79}},
  \bibinfo{pages}{235336} (\bibinfo{year}{2009}).

\bibitem{Jakobs2007}
\bibinfo{author}{Jakobs, S.~G.}, \bibinfo{author}{Meden, V.} \&
  \bibinfo{author}{Schoeller, H.}
\newblock \bibinfo{title}{{Nonequilibrium Functional Renormalization Group for
  Interacting Quantum Systems.}}
\newblock \emph{\bibinfo{journal}{Phys. Rev. Lett.}}
  \textbf{\bibinfo{volume}{99}}, \bibinfo{pages}{150603}
  (\bibinfo{year}{2007}).

\bibitem{Kennes2012}
\bibinfo{author}{Kennes, D.~M.}, \bibinfo{author}{Jakobs, S.~G.},
  \bibinfo{author}{Karrasch, C.} \& \bibinfo{author}{Meden, V.}
\newblock \bibinfo{title}{{Renormalization group approach to time-dependent
  transport through correlated quantum dots.}}
\newblock \emph{\bibinfo{journal}{Phys. Rev. B}} \textbf{\bibinfo{volume}{85}},
  \bibinfo{pages}{85113} (\bibinfo{year}{2012}).

\bibitem{Kennes2013}
\bibinfo{author}{Kennes, D.~M.}, \bibinfo{author}{Kashuba, O.},
  \bibinfo{author}{Pletyukhov, M.}, \bibinfo{author}{Schoeller, H.} \&
  \bibinfo{author}{Meden, V.}
\newblock \bibinfo{title}{{Oscillatory Dynamics and Non-Markovian Memory in
  Dissipative Quantum Systems.}}
\newblock \emph{\bibinfo{journal}{Phys. Rev. Lett.}}
  \textbf{\bibinfo{volume}{110}}, \bibinfo{pages}{100405}
  (\bibinfo{year}{2013}).

\bibitem{Muhlbacher2008}
\bibinfo{author}{M\"{u}hlbacher, L.} \& \bibinfo{author}{Rabani, E.}
\newblock \bibinfo{title}{{Real-Time Path Integral Approach to Nonequilibrium
  Many-Body Quantum Systems.}}
\newblock \emph{\bibinfo{journal}{Phys. Rev. Lett.}}
  \textbf{\bibinfo{volume}{100}}, \bibinfo{pages}{176403}
  (\bibinfo{year}{2008}).

\bibitem{Schiro2009}
\bibinfo{author}{Schir\'{o}, M.} \& \bibinfo{author}{Fabrizio, M.}
\newblock \bibinfo{title}{{Real-time diagrammatic Monte Carlo for
  nonequilibrium quantum transport.}}
\newblock \emph{\bibinfo{journal}{Phys. Rev. B}} \textbf{\bibinfo{volume}{79}},
  \bibinfo{pages}{153302} (\bibinfo{year}{2009}).

\bibitem{Werner2009}
\bibinfo{author}{Werner, P.}, \bibinfo{author}{Oka, T.} \&
  \bibinfo{author}{Millis, A.~J.}
\newblock \bibinfo{title}{{Diagrammatic Monte Carlo simulation of
  nonequilibrium systems.}}
\newblock \emph{\bibinfo{journal}{Phys. Rev. B}} \textbf{\bibinfo{volume}{79}},
  \bibinfo{pages}{35320} (\bibinfo{year}{2009}).

\bibitem{Nakajima1958}
\bibinfo{author}{Nakajima, S.}
\newblock \bibinfo{title}{{On Quantum Theory of Transport Phenomena.}}
\newblock \emph{\bibinfo{journal}{Progr. Theor. Phys.}}
  \textbf{\bibinfo{volume}{20}}, \bibinfo{pages}{948--959}
  (\bibinfo{year}{1958}).

\bibitem{Zwanzig1960}
\bibinfo{author}{Zwanzig, R.}
\newblock \bibinfo{title}{{Ensemble Method in the Theory of Irreversibility.}}
\newblock \emph{\bibinfo{journal}{J. Chem. Phys.}}
  \textbf{\bibinfo{volume}{33}}, \bibinfo{pages}{1338--1341}
  (\bibinfo{year}{1960}).

\bibitem{Tokuyama1976}
\bibinfo{author}{Tokuyama, M.} \& \bibinfo{author}{Mori, H.}
\newblock \bibinfo{title}{{Statistical-mechanical theory of random frequency
  modulations and generalized Brownian motions.}}
\newblock \emph{\bibinfo{journal}{Progr. Theor. Phys.}}
  \textbf{\bibinfo{volume}{55}}, \bibinfo{pages}{411--429}
  (\bibinfo{year}{1976}).

\bibitem{Shibata1977}
\bibinfo{author}{Shibata, F.}, \bibinfo{author}{Takahashi, Y.} \&
  \bibinfo{author}{Hashitsume, N.}
\newblock \bibinfo{title}{{A generalized stochastic liouville equation.
  Non-Markovian versus memoryless master equations.}}
\newblock \emph{\bibinfo{journal}{J. Stat. Phys.}}
  \textbf{\bibinfo{volume}{17}}, \bibinfo{pages}{171--187}
  (\bibinfo{year}{1977}).

\bibitem{Zwanzig2001}
\bibinfo{author}{Zwanzig, R.}
\newblock \emph{\bibinfo{title}{{Nonequilibrium statistical mechanics.}}}
  (\bibinfo{publisher}{Oxford University Press}, \bibinfo{city}{Oxford}, \bibinfo{year}{2001}).

\bibitem{Jin2008}
\bibinfo{author}{Jin, J.}, \bibinfo{author}{Zheng, X.} \& \bibinfo{author}{Yan,
  Y.}
\newblock \bibinfo{title}{{Exact dynamics of dissipative electronic systems and
  quantum transport: Hierarchical equations of motion approach.}}
\newblock \emph{\bibinfo{journal}{J. Chem. Phys.}}
  \textbf{\bibinfo{volume}{128}}, \bibinfo{pages}{234703}
  (\bibinfo{year}{2008}).

\bibitem{Cohen2011}
\bibinfo{author}{Cohen, G.} \& \bibinfo{author}{Rabani, E.}
\newblock \bibinfo{title}{{Memory effects in nonequilibrium quantum impurity
  models.}}
\newblock \emph{\bibinfo{journal}{Phys. Rev. B}} \textbf{\bibinfo{volume}{84}},
  \bibinfo{pages}{075150} (\bibinfo{year}{2011}).

\bibitem{Cohen2013a}
\bibinfo{author}{Cohen, G.}, \bibinfo{author}{Gull, E.},
  \bibinfo{author}{Reichman, D.~R.}, \bibinfo{author}{Millis, A.~J.} \&
  \bibinfo{author}{Rabani, E.}
\newblock \bibinfo{title}{{Numerically exact long-time magnetization dynamics
  at the nonequilibrium Kondo crossover of the Anderson impurity model.}}
\newblock \emph{\bibinfo{journal}{Phys. Rev. B}} \textbf{\bibinfo{volume}{87}},
  \bibinfo{pages}{195108} (\bibinfo{year}{2013}).

\bibitem{Pollard1996}
\bibinfo{author}{Pollard, W.~T.}, \bibinfo{author}{Felts, A.~K.} \&
  \bibinfo{author}{Friesner, R.~A.}
\newblock \emph{\bibinfo{title}{{The Redfield equation in condensed-phase
  quantum dynamics.}}}, vol. \bibinfo{volume}{XCIII} (\bibinfo{publisher}{John
  Wiley \& Sons, Inc.}, \bibinfo{city}{New York}, \bibinfo{year}{1996}).

\bibitem{Datta1995}
\bibinfo{author}{Datta, S.}
\newblock \emph{\bibinfo{title}{{Electronic transport in mesoscopic systems.}}}
  (\bibinfo{publisher}{Cambridge University Press}, \bibinfo{city}{Cambridge}, \bibinfo{year}{1995}).

\bibitem{Wangsness1953}
\bibinfo{author}{Wangsness, R.~K.} \& \bibinfo{author}{Bloch, F.}
\newblock \bibinfo{title}{{The Dynamical Theory of Nuclear Induction.}}
\newblock \emph{\bibinfo{journal}{Phys. Rev.}} \textbf{\bibinfo{volume}{89}},
  \bibinfo{pages}{728--739} (\bibinfo{year}{1953}).

\bibitem{Bloch1957}
\bibinfo{author}{Bloch, F.}
\newblock \bibinfo{title}{{Generalized theory of relaxation.}}
\newblock \emph{\bibinfo{journal}{Phys. Rev.}} \textbf{\bibinfo{volume}{411}},
  \bibinfo{pages}{1206--1222} (\bibinfo{year}{1957}).

\bibitem{Redfield1957}
\bibinfo{author}{Redfield, A.~G.}
\newblock \bibinfo{title}{{On the Theory of Relaxation Processes.}}
\newblock \emph{\bibinfo{journal}{IBM J. Res. Dev.}}
  \textbf{\bibinfo{volume}{1}}, \bibinfo{pages}{19--31} (\bibinfo{year}{1957}).

\bibitem{Scully1997}
\bibinfo{author}{Scully, M.~O.} \& \bibinfo{author}{Zubairy, M.~S.}
\newblock \emph{\bibinfo{title}{{Quantum optics.}}}
  (\bibinfo{publisher}{Cambridge University Press}, \bibinfo{year}{1997}).

\bibitem{Gardiner2004}
\bibinfo{author}{Gardiner, C.} \& \bibinfo{author}{Zoller, P.}
\newblock \emph{\bibinfo{title}{{Quantum noise.}}}, vol.~\bibinfo{volume}{56}
  (\bibinfo{publisher}{Springer}, \bibinfo{year}{2004}).

\bibitem{Breuer2002}
\bibinfo{author}{Breuer, H.-P.} \& \bibinfo{author}{Petruccione, F.}
\newblock \emph{\bibinfo{title}{{The theory of open quantum systems.}}}
  (\bibinfo{publisher}{Oxford University Press}, \bibinfo{year}{2002}).

\bibitem{Timm2008}
\bibinfo{author}{Timm, C.}
\newblock \bibinfo{title}{{Tunneling through molecules and quantum dots:
  Master-equation approaches.}}
\newblock \emph{\bibinfo{journal}{Phys. Rev. B}} \textbf{\bibinfo{volume}{77}},
  \bibinfo{pages}{195416} (\bibinfo{year}{2008}).

\bibitem{Koller2010}
\bibinfo{author}{Koller, S.}, \bibinfo{author}{Grifoni, M.},
  \bibinfo{author}{Leijnse, M.} \& \bibinfo{author}{Wegewijs, M.~R.}
\newblock \bibinfo{title}{{Density-operator approaches to transport through
  interacting quantum dots: Simplifications in fourth-order perturbation
  theory.}}
\newblock \emph{\bibinfo{journal}{Phys. Rev. B}} \textbf{\bibinfo{volume}{82}},
  \bibinfo{pages}{235307} (\bibinfo{year}{2010}).

\bibitem{Timm2011}
\bibinfo{author}{Timm, C.}
\newblock \bibinfo{title}{{Time-convolutionless master equation for quantum
  dots: Perturbative expansion to arbitrary order.}}
\newblock \emph{\bibinfo{journal}{Phys. Rev. B}} \textbf{\bibinfo{volume}{83}},
  \bibinfo{pages}{115416} (\bibinfo{year}{2011}).

\bibitem{Cohen2013}
\bibinfo{author}{Cohen, G.}, \bibinfo{author}{Wilner, E.~Y.} \&
  \bibinfo{author}{Rabani, E.}
\newblock \bibinfo{title}{{Generalized projected dynamics for non-system
  observables of non-equilibrium quantum impurity models.}}
\newblock \emph{\bibinfo{journal}{New J. Phys.}} \textbf{\bibinfo{volume}{15}},
  \bibinfo{pages}{073018} (\bibinfo{year}{2013}).
  
\bibitem{Dirac1927}
\bibinfo{author}{Dirac, P. A.~M.}
\newblock \bibinfo{title}{{The Quantum Theory of the Emission and Absorption of
  Radiation.}}
\newblock \emph{\bibinfo{journal}{Proc. R. Soc. A}}
  \textbf{\bibinfo{volume}{114}}, \bibinfo{pages}{243--265}
  (\bibinfo{year}{1927}).

\bibitem{Alicki1977}
\bibinfo{author}{Alicki, R.}
\newblock \bibinfo{title}{{The Markov master equations and the Fermi golden
  rule.}}
\newblock \emph{\bibinfo{journal}{Int. J. Theor. Phys.}}
  \textbf{\bibinfo{volume}{16}}, \bibinfo{pages}{351--355}
  (\bibinfo{year}{1977}).
  
\bibitem{Bagrets2003}
\bibinfo{author}{Bagrets, D.~A.} \& \bibinfo{author}{Nazarov, Y.~V.}
\newblock \bibinfo{title}{{Full counting statistics of charge transfer in
  Coulomb blockade systems.}}
\newblock \emph{\bibinfo{journal}{Phys. Rev. B}} \textbf{\bibinfo{volume}{67}},
  \bibinfo{pages}{085316} (\bibinfo{year}{2003}).
  
\bibitem{Braggio2006}
\bibinfo{author}{Braggio, A.}, \bibinfo{author}{K\"{o}nig, J.} \&
  \bibinfo{author}{Fazio, R.}
\newblock \bibinfo{title}{{Full Counting Statistics in Strongly Interacting
  Systems: Non-Markovian Effects.}}
\newblock \emph{\bibinfo{journal}{Phys. Rev. Lett.}}
  \textbf{\bibinfo{volume}{96}}, \bibinfo{pages}{026805}
  (\bibinfo{year}{2006}).
  
\bibitem{Esposito2009}
\bibinfo{author}{Esposito, M.}
\newblock \bibinfo{title}{{Nonequilibrium fluctuations, fluctuation theorems,
  and counting statistics in quantum systems.}}
\newblock \emph{\bibinfo{journal}{Rev. Mod. Phys.}}
  \textbf{\bibinfo{volume}{81}}, \bibinfo{pages}{1665--1702}
  (\bibinfo{year}{2009}).

\bibitem{Gorini1976}
\bibinfo{author}{Gorini, V.}, \bibinfo{author}{Kossakowski, A.} \&
  \bibinfo{author}{Sudarshan, E. C.~G.}
\newblock \bibinfo{title}{{Completely positive dynamical semigroups of N-level
  systems.}}
\newblock \emph{\bibinfo{journal}{J. Math. Phys.}}
  \textbf{\bibinfo{volume}{17}}, \bibinfo{pages}{821--825}
  (\bibinfo{year}{1976}).

\bibitem{Lindblad1976}
\bibinfo{author}{Lindblad, G.}
\newblock \bibinfo{title}{{On the generators of quantum dynamical semigroups.}}
\newblock \emph{\bibinfo{journal}{Commun. Math. Phys.}}
  \textbf{\bibinfo{volume}{48}}, \bibinfo{pages}{119--130}
  (\bibinfo{year}{1976}).

\bibitem{Alicki2007}
\bibinfo{author}{Alicki, R.} \& \bibinfo{author}{Lendi, K.}
\newblock \emph{\bibinfo{title}{{Quantum Dynamical Semigroups and
  Applications.}}} (\bibinfo{publisher}{Springer}, \bibinfo{year}{2007}).

\bibitem{Prosen2008}
\bibinfo{author}{Prosen, T.}
\newblock \bibinfo{title}{{Third quantization: a general method to solve master
  equations for quadratic open Fermi systems.}}
\newblock \emph{\bibinfo{journal}{New J. Phys.}} \textbf{\bibinfo{volume}{10}},
  \bibinfo{pages}{043026} (\bibinfo{year}{2008}).

\bibitem{Dzhioev2011}
\bibinfo{author}{Dzhioev, A.~A.} \& \bibinfo{author}{Kosov, D.~S.}
\newblock \bibinfo{title}{{Super-fermion representation of quantum kinetic
  equations for the electron transport problem.}}
\newblock \emph{\bibinfo{journal}{J. Chem. Phys.}}
  \textbf{\bibinfo{volume}{134}}, \bibinfo{pages}{044121}
  (\bibinfo{year}{2011}).

\bibitem{Horstmann2013}
\bibinfo{author}{Horstmann, B.}, \bibinfo{author}{Cirac, J.~I.} \&
  \bibinfo{author}{Giedke, G.}
\newblock \bibinfo{title}{{Noise-driven dynamics and phase transitions in
  fermionic systems.}}
\newblock \emph{\bibinfo{journal}{Phys. Rev. A}} \textbf{\bibinfo{volume}{87}},
  \bibinfo{pages}{012108} (\bibinfo{year}{2013}).

\bibitem{Mitchell2010}
\bibinfo{author}{Mitchell, A.~K.} \& \bibinfo{author}{Logan, D.~E.}
\newblock \bibinfo{title}{{Two-channel Kondo phases and frustration-induced
  transitions in triple quantum dots.}}
\newblock \emph{\bibinfo{journal}{Phys. Rev. B}} \textbf{\bibinfo{volume}{81}},
  \bibinfo{pages}{75126} (\bibinfo{year}{2010}).

\bibitem{Mitchell2013}
\bibinfo{author}{Mitchell, A.~K.}, \bibinfo{author}{Jarrold, T.~F.},
  \bibinfo{author}{Galpin, M.~R.} \& \bibinfo{author}{Logan, D.~E.}
\newblock \bibinfo{title}{{Local moment formation and Kondo screening in
  impurity trimers.}}
\newblock \emph{\bibinfo{journal}{J. Phys. Chem. B}}
  \textbf{\bibinfo{volume}{117}}, \bibinfo{pages}{12777--86}
  (\bibinfo{year}{2013}).

\bibitem{Kumagai2012}
\bibinfo{author}{Kumagai, T.} \emph{et~al.}
\newblock \bibinfo{title}{{H-atom relay reactions in real space.}}
\newblock \emph{\bibinfo{journal}{Nat. Mater.}} \textbf{\bibinfo{volume}{11}},
  \bibinfo{pages}{167--172} (\bibinfo{year}{2012}).

\bibitem{Frederiksen2014}
\bibinfo{author}{Frederiksen, T.}, \bibinfo{author}{Paulsson, M.} \&
  \bibinfo{author}{Ueba, H.}
\newblock \bibinfo{title}{{Theory of action spectroscopy for single-molecule
  reactions induced by vibrational excitations with STM.}}
\newblock \emph{\bibinfo{journal}{Phys. Rev. B}} \textbf{\bibinfo{volume}{89}},
  \bibinfo{pages}{35427} (\bibinfo{year}{2014}).

\bibitem{Zimmermann2011}
\bibinfo{author}{Zimmermann, B.}, \bibinfo{author}{M\"{u}ller, T.},
  \bibinfo{author}{Meineke, J.}, \bibinfo{author}{Esslinger, T.} \&
  \bibinfo{author}{Moritz, H.}
\newblock \bibinfo{title}{{High-resolution imaging of ultracold fermions in
  microscopically tailored optical potentials.}}
\newblock \emph{\bibinfo{journal}{New J. Phys.}} \textbf{\bibinfo{volume}{13}},
  \bibinfo{pages}{043007} (\bibinfo{year}{2011}).
  
\bibitem{Li2014}
\bibinfo{author}{Li, G.}, \bibinfo{author}{Antipov, A.~E.},
  \bibinfo{author}{Rubtsov, A.~N.}, \bibinfo{author}{Kirchner, S.} \&
  \bibinfo{author}{Hanke, W.}
\newblock \bibinfo{title}{{Competing phases of the Hubbard model on a
  triangular lattice: Insights from the entropy.}}
\newblock \emph{\bibinfo{journal}{Phys. Rev. B}} \textbf{\bibinfo{volume}{89}},
  \bibinfo{pages}{161118} (\bibinfo{year}{2014}).

\bibitem{Reiter2012}
\bibinfo{author}{Reiter, F.} \& \bibinfo{author}{S\o rensen, A.~S.}
\newblock \bibinfo{title}{{Effective operator formalism for open quantum
  systems.}}
\newblock \emph{\bibinfo{journal}{Phys. Rev. A}} \textbf{\bibinfo{volume}{85}},
  \bibinfo{pages}{032111} (\bibinfo{year}{2012}).

\bibitem{Schuetz2013}
\bibinfo{author}{Schuetz, M.}, \bibinfo{author}{Kessler, E.},
  \bibinfo{author}{Vandersypen, L.}, \bibinfo{author}{Cirac, J.} \&
  \bibinfo{author}{Giedke, G.}
\newblock \bibinfo{title}{{Steady-State Entanglement in the Nuclear Spin
  Dynamics of a Double Quantum Dot.}}
\newblock \emph{\bibinfo{journal}{Phys. Rev. Lett.}}
  \textbf{\bibinfo{volume}{111}}, \bibinfo{pages}{246802}
  (\bibinfo{year}{2013}).
  
\bibitem{Cramer2013}
\bibinfo{author}{Cramer, C.~J.}
\newblock \emph{\bibinfo{title}{{Essentials of computational chemistry:
  theories and models.}}} (\bibinfo{publisher}{John Wiley \& Sons},
  \bibinfo{year}{2013}).

\end{thebibliography}
\end{document}